\documentclass[preprint,12pt]{elsarticle}

\usepackage{graphicx}
\usepackage{amsmath}
\usepackage{amssymb}
\usepackage{url}
\usepackage[dvipsnames]{color}

\usepackage{yhmath}
\usepackage[ruled,vlined]{algorithm2e}

\usepackage{srctex} 

\def\m#1{\mathsf{#1}} 

\newcommand{\cl}[2]{\ensuremath{\mathit{Cl}_{#1,#2}}}

\newcommand{\bbR}{\ensuremath{\mathbb{R}}}
\newcommand{\bbC}{\ensuremath{\mathbb{C}}}
\newcommand{\bbH}{\ensuremath{\mathbb{H}}}

\newcommand{\cliffordconjugate}[1]{\widetilde{\wideparen{#1}}}

\def\A{\mathsf{A}}
\def\B{\mathsf{B}}

\def\m#1{\mathsf{#1}}

\def\e#1{\mathbf{e}_{#1}} 

\newcommand{\ba}{\ensuremath{\mathbf{a}}}
\newcommand{\bb}{\ensuremath{\mathbf{b}}}

\newcommand{\bv}{\ensuremath{\mathbf{v}}}
\newcommand{\bV}{\ensuremath{\mathbf{V}}}

\newcommand{\cA}{\ensuremath{\mathcal{A}}}
\newcommand{\cB}{\ensuremath{\mathcal{B}}}

\newcommand{\red}[1]{\textcolor{red}{#1}}
\newcommand{\blue}[1]{\textcolor{blue}{#1}}

\begin{document}

\begin{frontmatter}



\title{Square root of a multivector of Clifford algebras in~3D: A game with signs}

\author{A.~Acus}
\address{Institute of Theoretical Physics and Astronomy,\break
Vilnius University,\break Saul{\.e}tekio 3, LT-10257 Vilnius,
Lithuania, arturas.acus@tfai.vu.lt}

\author{A.~Dargys}

\address{%
Center for Physical Sciences and Technology,\break Semiconductor
Physics Institute,\break Saul{\.e}tekio 3, LT-10257 Vilnius,
Lithuania, adolfas.dargys@ftmc.lt}

\begin{abstract}
An algorithm to extract the square root from a multivector (MV) in
real Clifford algebras $\cl{p}{q}$, where $n=p+q\le 3$, in
radicals is presented. It is shown that  in  \cl{3}{0},  \cl{1}{2}
and \cl{0}{3} algebras there are up to four isolated square roots
in a case of the most general (generic) MV. The algebra \cl{2}{1}
is an exception and there the MV can have up to 16 isolated roots.
In addition, a continuum of roots has been found in all Clifford
algebras except $p+q=1$. Examples which clarify algorithm are
provided to illustrate the properties of roots in all  $n=3$
algebras. The results may be useful in solving nonlinear
equations, for example Clifford-Riccati equation.
\end{abstract}



\end{frontmatter}


\section{Introduction}\label{sec:intro}
The square root has a long history. Solution by radicals of the
cubic equation was first published  in 1545 by G.~Cardano.
Simultaneously, a concept of square root of a negative number  has
been developed~\cite{Grant2015}. In 1872 A.~Cayley was the first
to carry over the  square root to matrices~\cite{Cayley1872}. In
the recent book by  N.~J. Higham~\cite{Higham08}, where an
extensive literature is presented on nonlinear functions of
matrices, two sections are devoted to matrix square roots. In the context of Clifford
algebra (CA) the main attention up till now was concentrated on
the square roots of quaternions~\cite{Niven1942,Opfer2017}, or
their derivatives such as coquaternions (also called split
quaternions), or
nectarines~\cite{Falcao2018,Opfer2017,Ozdemir2009}. The square
root of biquaternion (complex quaternion) was considered
in~\cite{Sangwine2006}. The quaternions and related objects are
isomorphic to one of $n=2$ algebras \cl{0}{2}, \cl{1}{1},
\cl{2}{0} and, therefore, the quaternionic square root analysis
can be easily rewritten in  terms of CA (see~\ref{appendix:dim2}).
In this paper we shall mainly be interested in higher, namely,
$n=3$ Clifford algebras (CAs), where the main object is the
8-component MV.

For CAs of dimension $n\ge 3$ the investigation and understanding
of square root properties is still in infancy. The
most akin to the present paper are the investigation of conditions
for existence of square root of
$-1$~\cite{Sangwine2006,Hitzer2011,Hitzer2013}. The existence of
such roots allows to extend the  Fourier transform to MVs, where
they are used in formulating Clifford-Fourier transform and CA
based wavelet theories~\cite{Hitzer13B}.

Our preliminary investigation~\cite{Dargys2019} on this subject
was concerned with square roots of individual MV grades such as
scalar, vector, bivector, pseudoscalar, or their simple
combinations. For this purpose we have applied the Gr{\"o}bner
basis algorithm to analyze the  system of nonlinear polynomial
equations that ensue from the MV equation $\A^2=\B$, where $\A$
and $\B$ are the MVs. The Gr{\"o}bner basis is accessible in
symbolic mathematical packages such as \textit{Mathematica} and
\textit{Maple}. Specifically, the \textit{Mathematica} commands
such as \textbf{Reduce[~]}, \textbf{Solve[~]},
\textbf{Eliminate[~]} and others also employ the Gr{\"o}bner basis
to solve nonlinear problems. With the help of them we were able to
find new properties of roots for $n=3$ case, namely, that the MVs
may have no roots, a single or multiple isolated roots, or even an
infinite number (continuum) of roots in 4D parameter spaces or
smaller dimensions.

In this paper we continue our~\cite{Dargys2019} investigations of
the square root problem in real CAs for $n=3$ case. In particular,
we examine and provide explicit conditions for a MV to have
discrete and continuum of roots, and how to express real
root coefficients in radicals. For this purpose a symbolic
package based on {\it Mathematica} system was
written~\cite{AcusDargys2023} that appeared to be invaluable
both for detecting specific solutions of the nonlinear CA
equation $\A^2=\B$ and for numerical checks in general.

In Sec.~\ref{sec:notations}  the notation is introduced. The
algorithm to calculate the square root of a generic MV and special
cases that follow are given in Secs.~\ref{sec:Cl30}-\ref{sec:Cl21}
for $\cl{3}{0}\simeq \cl{1}{2}$, \cl{0}{3}, and \cl{2}{1}
algebras, respectively. The algorithm is illustrated by a number
of examples. The conclusions are drawn
in~Sec.~\ref{sec:conclusions}. For completeness, in
\ref{appendix:dim1} and \ref{appendix:dim2} the MV square roots
are  presented for lower dimensional CAs.

\section{Notation}\label{sec:notations}
For $n=3$,  general MV can be expanded in the orthonormal basis
that consists of $2^{n}=8$ elements listed in inverse degree
lexicographic ordering,\footnote{\label{note1} Note an increasing order of digits
in indices. Therefore we write $\e{13}$ instead of
$\e{31}=-\e{13}$. This convention is reflected in opposite signs
of some terms in formulas.}
\begin{equation}\label{basisDim3}
\{1,\e{1},\e{2},\e{3},\e{12},\e{13},\e{23},\e{123}\equiv I\},
\end{equation}
where $\e{i}$ are basis vectors and $\e{ij}$ are the bivectors
(oriented planes). The last term is the pseudoscalar. The number
of subscripts indicates the grade of basis element. The scalar is
a grade-0 element, the vectors $\e{i}$ are the grade-1 elements
etc. In the orthonormalized basis the geometric (Clifford)
products of basis vectors satisfy the anticommutation relation,
 \begin{equation}\label{anticom}
 \e{i}\e{j}+\e{j}\e{i}=\pm 2\delta_{ij}.
 \end{equation}
For \cl{3}{0} and \cl{0}{3} algebras the squares of basis vectors,
correspondingly, are $\e{i}^2=+1$ and $\e{i}^2=-1$, where
$i=1,2,3$. For mixed signature algebras such as  \cl{2}{1} and
\cl{1}{2} we have $\e{1}^2=\e{2}^2=1$, $\e{3}^2=-1$ and
$\e{1}^2=1$, $\e{2}^2=\e{3}^2=-1$, respectively. The sign of
squares of higher grade elements is determined by squares of
vectors and the property~\eqref{anticom}. For example, in
\cl{3}{0} we have
$\e{12}^2=\e{12}\e{12}=-\e{1}\e{2}\e{2}\e{1}=-\e{1}(+1)\e{1}=-\e{1}\e{1}=-1$.
However, in \cl{1}{2} similar computation gives
$\e{12}^2=-\e{1}\e{2}\e{2}\e{1}=-\e{1}(-1)\e{1}=\e{1}\e{1}=+1$.

When $n=3$, a MV $\A$ in real CA can be expanded in the
basis~\eqref{basisDim3},
\begin{equation}\begin{split}\label{mvAA}
\A=&a_0+a_1\e{1}+a_2\e{2}+a_3\e{3}+a_{12}\e{12}+a_{23}\e{23}+a_{13}\e{13}+a_{123}I\\
\equiv&a_0+\ba+\cA+a_{123}I,
\end{split}\end{equation} where $a_i$, $a_{ij}$ and $a_{123}$ are the real
coefficients, and $\ba=a_1\e{1}+a_2\e{2}+a_3\e{3}$ and
$\cA=a_{12}\e{12}+a_{23}\e{23}+a_{13}\e{13}$ is, respectively, the
vector and the bivector. We will seek for a real MV $\A$
(with real coefficients), the square of which satisfies
\begin{equation}\label{AA}
  \A\A\equiv\A^2=\B=b_0+\mathbf{b}+\cB+b_{123}I.
\end{equation}
The MV $\A$ is called  a square root of $\B$. In Eq.~\eqref{AA}
the square $\A^2$ has been expanded in the orthonormal basis where
$b_0,\mathbf{b},\cB$ and $I\equiv I_3 $ denote, respectively, a
scalar, a vector ($\bb=b_1\e{1}+b_2\e{2}+b_3\e{3}$), a bivector
($\cB=b_{12}\e{12}+b_{23}\e{23}+b_{13}\e{13}$) and a pseudoscalar.
The representation~\eqref{mvAA} is not convenient for our problem,
therefore, for all 3D algebras \cl{3}{0}, \cl{0}{3}, \cl{1}{2},
and \cl{2}{1} a more symmetric representation is introduced,
\begin{equation}
\A=s+\bv+(S+\bV)I,\label{mvA30}
\end{equation}
where now both $s$ and $S$ are the real scalars and both
$\bv=v_1\e{1}+v_2\e{2}+v_3\e{3}$ and
$\bV=V_1\e{1}+V_2\e{2}+V_3\e{3}$ are the vectors with real
coefficients $v_i$ and $V_i$. The MV representation~\eqref{mvA30}
allows to disentangle the coupled nonlinear equations in a regular
manner for all listed algebras. To select the scalar $s$
in~\eqref{mvA30}, the grade selector
$\langle\m{A}\rangle\equiv\langle\m{A}\rangle_0=s$ is used. The
pseudoscalar part  can be extracted by
$\langle\m{A}\rangle\equiv\langle-\m{A}I\rangle_0=S$, and
similarly for other grades. More about CAs and MV properties can
be found, for example, in books~\cite{Doran03,Lounesto97}.

When $n=1,2$ the MV square root algorithm simplifies
substantially. All needed formulas are presented in the
\ref{appendix:dim1} and \ref{appendix:dim2}, respectively.

\section{Square roots in \cl{3}{0} and \cl{1}{2} algebras\label{sec:Cl30}}
This section describes the method of substitution of variables in
CA which paves a direct way to square root algorithm (also
see the \ref{appendix:algorithm}). The Euclidean \cl{3}{0}
algebra is the most simple one among $n=3$ algebras. The algebra
\cl{1}{2} is isomorphic to \cl{3}{0}, therefore the algorithm for
this algebra follows the same route except that there some
notational differences appear.

The goal is to solve nonlinear MV equation
$\A^2=\B$, where $\B= b_0+b_1 \e{2}+b_2\e{2}+b_3\e{3}+
b_{12}\e{23}+b_{13}\e{13}+b_{23} \e{23}+b_{123} I$ and  $\A$ is
unknown. The latter  may have the general form~\eqref{mvA30}.
Expanding $\A^2$ in components and equating (real) coefficients at
basis elements to respective coefficients in $\B$ one obtains a
system of eight nonlinear equations:
\begin{align}
b_0&=s^2-S^2+\bv^2-\bV^2, &b_{123}&=2(sS+\bv\cdot\bV)\label{MVM1Cl30}, \\
b_1&=2(s v_1-S V_1), &b_{23}&=2(s V_1+S v_1)\label{MVM2Cl30}, \\
b_2&=2(s v_2-S V_2), &b_{13}&=2(s V_2+S v_2)\label{MVM3Cl30}, \\
b_3&=2(s v_3-S V_3), &b_{12}&=2(s V_3+S v_3)\label{MVM4Cl30}.
\end{align}
For all algebras the square root algorithm splits into two cases:
the \textit{generic} case where either $s^2+S^2\neq 0$ (in
\cl{3}{0} and \cl{1}{2}) or $s^2-S^2\neq 0$ (in \cl{0}{3} and
\cl{2}{1}), and the \textit{special} case where $s^2+S^2= 0$ (in
\cl{3}{0} and \cl{1}{2}) or $s^2-S^2= 0$ (in \cl{0}{3} and
\cl{2}{1}).

\subsection{The generic case $s^2+S^2\neq 0$}
The system of six Eqs. \eqref{MVM2Cl30}-\eqref{MVM4Cl30} is
linear in new variables $v_i$ and $V_i$ in~\eqref{mvA30}. It has
very simple solution which is a key to analysis that
follows\footnote{Note, the symmetry of Eqs.~\eqref{vV1Cl30}
and~\eqref{vV2Cl30} with respect to pairs ($v_2,V_2$), ($v_1,V_1$)
and ($v_3,V_3$) differ as explained in Footnote~\ref{note1}. It
can be restored if $b_{13}$ in is replaced by $-b_{31}$.},
\begin{align}
v_1&=\frac{b_1s+b_{23}S}{2(s^2+S^2)},
 &v_2&=\frac{b_{2}s-b_{13}S}{2(s^2+S^2)},
  &v_3&=\frac{b_3s+b_{12}S}{2(s^2+S^2)}, \label{vV1Cl30} \\
V_1&=\frac{b_{23}s-b_{1}S}{2(s^2+S^2)},
&V_2&=-\frac{b_{13}s+b_{2}S}{2(s^2+S^2)},
&V_3&=\frac{b_{12}s-b_{3}S}{2(s^2+S^2)}.\label{vV2Cl30}
\end{align}
The Eqs. \eqref{vV1Cl30}-\eqref{vV2Cl30} express the components of
vectors $\bv$ and $\bV$ in terms of scalars $s$ and $S$, which are
to be determined from a pair of equations~\eqref{MVM1Cl30}. The
solution is valid when $s^2+S^2\neq 0$, i.e., when either $s\neq
0$ or $S\neq 0$, or both $s$ and $S$ are nonzero scalars. If these
conditions are not satisfied we have the subcase $s=S=0$.  After
substitution of \eqref{vV1Cl30}--\eqref{vV2Cl30}, i.e., of
$(v_1,v_2,v_3)$ and $(V_1,V_2,V_3)$, into~\eqref{MVM1Cl30} we get
a system of two coupled algebraic equations for two unknowns $s$
and $S$,
\begin{equation}\begin{split}\label{sysACl30}
4(b_0-s^2+S^2)(s^2+S^2)^2=&+(b_1s+b_{23}S)^2+(b_2s-b_{13}S)^2+(b_3s+b_{12}S)^2\\
  &-(b_{23}s-b_1S)^2-(b_{13}s+b_2S)^2-(b_{12}s-b_3S)^2,\\
  2(b_{123}-2sS)(s^2+S^2)^2=&+(b_1s+b_{23}S)(b_{23}s-b_{1}S)-(b_{2}s-b_{13}S)\\
&\times(b_{13}s+b_{2}S)+(b_3s+b_{12}S)(b_{12}s-b_3S).
 \end{split}\end{equation}
The system~\eqref{sysACl30} has exactly four solutions that can be
expressed in radicals. If  new variables $t$ and $T$ are
introduced and substitution
\begin{equation}\label{sysASCl30}
  s\,S = t,\qquad  \tfrac{1}{2}(-s^2 + S^2) = T,
\end{equation}
is used  the system~\eqref{sysACl30} reduces  to
\begin{equation}\label{sysARCl30}
\begin{split}
&(b_{0}+4 T) (4 t-b_{123})-b_{I}/2=0,\\
&b_{S}-(b_{0}-b_{123}+4 T+4 t) (b_{0}+b_{123}+4 T-4 t)=0.
\end{split}\end{equation}
In~\eqref{sysARCl30}, coordinate-free abbreviations $b_{S}$
and $b_{I}$ have been introduced,
\begin{equation}\label{bSbICl30}
\begin{split}
&b_{S}= \langle \B \cliffordconjugate{\B}\rangle_0 =
b_{0}^2-b_{1}^2-b_{2}^2-b_{3}^2+b_{12}^2+b_{13}^2+b_{23}^2-b_{123}^2,\\
&b_{I}= \langle \B \cliffordconjugate{\B} I \rangle_0 = 2 b_{3}
b_{12}-2 b_{2} b_{13}+2 b_{1} b_{23}-2 b_{0} b_{123}.
\end{split}
\end{equation}
In \eqref{bSbICl30} the MV $\cliffordconjugate{\B}$ denotes the
Clifford conjugate of $\B$, where tilde is the grade
reversion and cap is the grade inversion. Note, that for
remaining algebras \cl{0}{3}, \cl{2}{1}, and \cl{2}{1} the signs
of individual terms inside $b_S$ and $b_I $ all are different. As
we shall see below, the square  roots for all $n=3$ algebras are
predetermined by four real  coefficients only, namely, $b_0$,
$b_{123}$, $b_{S}$, and~$b_{I}$.

After  substitution of~\eqref{sysASCl30}, the resulting system of
equations~\eqref{sysARCl30} are of degree $\le 4$. Thus, we
conclude that the initial system \eqref{sysACl30} can be
explicitly solved in radicals. In particular, two real solutions
of \eqref{sysASCl30} have the form
\begin{equation}\label{solACl30}
\begin{split}
  &\Bigl(s_{1,2} =\pm\sqrt{-T+\sqrt{T^2+ t^2}},\qquad S_{1,2}
  =\pm\frac{t}{\sqrt{-T+\sqrt{T^2+ t^2}}}\Bigr),
\end{split}\end{equation}
where the signs in pairs $(s_i,S_i)$ must be identical,
plus or minus. The denominator of $S_{1,2}$ becomes zero if
$s=S=0$. The remaining two solutions of \eqref{sysASCl30}, which
can be obtained from \eqref{solACl30} by the substitution
$\sqrt{T^2+ t^2}\to -\sqrt{T^2+ t^2}$, are complex valued due to
the inequality $\sqrt{T^2+ t^2}\ge T$ and therefore must be
rejected.

The two real-valued solutions of Eq.~\eqref{sysARCl30} are
\begin{equation}\label{solBCl30}
\begin{cases}
  \Bigl(  t_{1,2}=\frac{1}{4} \Bigl(b_{123} \pm \frac{1}{\sqrt{2}}\sqrt{-b_{S}+\sqrt{D}}\Bigr),\
  T_{1,2}=\frac{1}{4}\Bigl(\frac{\pm b_{I}}{\sqrt{2}\sqrt{-b_{S}+\sqrt{D}}}-b_{0}\Bigr)\Bigr),\\
  \hphantom{t_{1,2}=\frac{1}{4} \Bigl(b_{123} \pm \frac{1}{\sqrt{2}}\sqrt{-b_{S}+\sqrt{D}}\Bigr),\
  T_{1,2}=\frac{1}{4}b_{123}\,}
  \quad \textrm{if}\quad -b_{S}+\sqrt{D}>0,
  \\[2pt]
  \bigl( t_{1,2}= b_{123}/{4}, \
    T_{1,2}= \frac{1}{4} \bigl(\pm\sqrt{b_{S}}-b_{0}\bigr)\bigr),\ \textrm{if}\ -b_{S}+\sqrt{D}=0\ \&\
    b_{S}>0.
\end{cases}
\end{equation}
No additional conditions  are required  for the determinant
$D=b_{S}^2+b_{I}^2 \ge 0$ of the MV $\m{B}$, since for
\cl{3}{0} algebra it is always positive definite $D\ge b_{S}$
(refer to \cite{Acus2018,Acus2022} how to compute MV
determinant). Again, we should take the same signs for $t_i$ and
$T_i$. The two complex valued solutions of \eqref{sysARCl30},
which can be obtained from \eqref{solBCl30} by substitution
$\sqrt{D}\to -\sqrt{D}$, must be rejected. The denominator of
$T_{1,2}$ in \eqref{solBCl30} turns into zero when
$b_{S}=\sqrt{D}$, i.e., when $b_{I}=0$.

To summarize, starting from \eqref{solBCl30} and then going to
\eqref{solACl30}, and finally to formulas
\eqref{vV1Cl30}-\eqref{vV2Cl30}, one obtains four explicit real
solutions which completely determine the square root
  $\A=\sqrt{\B}$ of generic MV $\B$ in terms of radicals
$\A=s+\bv+(S+\bV)I$ of real Clifford algebra $\cl{3}{0}$.

\subsection{The special case $s^2+S^2= 0$\label{subsec:Cl30spec}}
The only special case in \cl{3}{0} corresponds to $s=S=0$.
In the subcases $s = S\neq 0$ and $s = -S\neq 0$ one can
rewrite expressions \eqref{solACl30} in a simpler form. In
particular, when
  $s=S\neq 0$ we have
\begin{equation}\begin{split}\label{solSspec1Cl30}
s_{1,2} =&\begin{cases}
\pm\tfrac12\sqrt{b_{123}+\dfrac{b_I}{2 b_0}} & \textrm{if}\quad b_0\neq  0,\\
\pm\tfrac12\sqrt{b_{123}\pm\sqrt{-b_S}} & \textrm{if}\quad b_0 =  0,
  \end{cases}
\end{split}\end{equation}
and when $s = -S\neq 0$
\begin{equation}\begin{split}\label{solSspec2}
s_{1,2} =&\begin{cases}
\pm\tfrac12\sqrt{-b_{123}-\dfrac{b_I}{2 b_0}} & \textrm{if}\quad b_0\neq  0,\\
\pm\tfrac12\sqrt{-b_{123}\pm\sqrt{-b_S}} & \textrm{if}\quad b_0 =  0,
  \end{cases}
\end{split}\end{equation}
where all expressions inside square roots are assumed to be
positive.

The case $s = S = 0$ is special, because the  condition implies
that the number of square roots of $\B$ may be
infinite.\footnote{The case of simple MV roots is given
in~\cite{Dargys2019}.} Indeed, in this case the Eqs.
\eqref{MVM2Cl30}-\eqref{MVM4Cl30} are compatible only if the
vector $(b_1, b_2, b_3)$ and bivector $(b_{12}, b_{13}, b_{23})$
coefficients  are  zeros. Then, the Eq.~\eqref{MVM1Cl30} reduces
to
\begin{equation}\label{solSspec3Cl30etro}
b_0=\bv^2-\bV^2,\quad b_{123}=2(\bv\cdot\bV),
\end{equation}
where $\bv^2=v_1^2+v_2^2+v_3^2$ and $\bV^2=V_1^2+V_2^2+V_3^2$  for
\cl{3}{0}. Since, in general, we have $3+3=6$ unknowns which must
satisfy Eqs.~\eqref{solSspec3Cl30etro} we are left with four real
arbitrary parameters as will be explicitly demonstrated in
\textit{Example}~1. The solution therefore makes a four
dimensional (or smaller) set of real-valued MV coefficients.
It is interesting  that the both expressions
in~\eqref{solSspec3Cl30etro} have very clear geometric
interpretation. Indeed, if the ends of vectors $\bv$ and $\bV$
represent two concentric spheres then the coefficient $b_0$
controls the lengths of radii $|\bv|$ and $|\bV|$, while the
pseudoscalar  coefficient $b_{123}$  controls the angle between
the vectors $\bv$ and~$\bV$. From this follows that, due to
periodicity of the angle,  one can introduce principal value for
coefficient $b_{123}$. Similar property, i.e., the multiplicity of
roots and the existence of principal angle in a complex plane are
well-known in case complex numbers~\cite{Korn-Korn1961}.

The diagram of the described algorithm  is presented
in~\ref{appendix:algorithm}.

\subsection{$\cl{1}{2}\simeq\cl{3}{0}$ algebra}\label{sec:Cl12}

In paper~\cite{Marchuk2015} it is shown that ``...for odd
$n \ge 3$, there are three classes of isomorphic Clifford algebras
what is consistent with Cartan's classification of real Clifford
algebras.'' In particular, two algebras, \cl{3}{0} and \cl{1}{2},
are represented by $2\times 2$ complex matrices $\bbC(2)$.
The similarity between square root expressions obtained
below  also confirms that these two algebras fall into the same
isomorphism class. On the other hand, the algebras \cl{0}{3} and
\cl{2}{1} are represented by blocked $2\times 2$ and $1\times 1$
matrices, respectively ${^2}\bbR(2)$ and ${^2}\bbH(1)$. Therefore,
they belong to different classes. Indeed, as we shall show later,
the analysis of roots in \cl{2}{1} is only roughly similar
to that in \cl{0}{3}. However,  between \cl{2}{1}
and \cl{0}{3} there  are distinctions: they are isomorphic to
different, real and quaternionic  matrices.

As far as \cl{1}{2} algebra  is concerned, its difference from
\cl{3}{0} is contained in the explicit expression for $b_{S}$,
\begin{align}
&b_{S}=\langle \m{B} \cliffordconjugate{\m{B}}\rangle_0 =
b_{0}^2-b_{1}^2+b_{2}^2+b_{3}^2-b_{12}^2-b_{13}^2+b_{23}^2-b_{123}^2,
\\
&b_{I}=\langle \m{B} \cliffordconjugate{\m{B}} I \rangle_0=
  2 b_{3} b_{12}-2 b_{2} b_{13}+2 b_{1} b_{23}-2 b_{0} b_{123},  \\
  &D=b_{S}^2+b_{I}^2,\qquad\qquad \ \cl{1}{2}
\end{align}
and expressions for $v_i$ and $V_i$
\begin{align}
v_1&=\frac{b_1s+b_{23}S}{2(s^2+S^2)},
   &v_2&=\frac{b_{2}s+b_{13}S}{2(s^2+S^2)},
    &v_3&=\frac{b_3 s -b_{12}S}{2(s^2+S^2)}, \label{vV1Cl12} \\
V_1&=\frac{b_{23}s-b_{1}S}{2(s^2+S^2)},
&V_2&=\frac{b_{13}S-b_{2}S}{2(s^2+S^2)},
&V_3&=-\frac{b_{12}s+b_{3}S}{2(s^2+S^2)}.\label{vV2Cl12}
\end{align}
The expressions for $b_{I}$ and $D$ (the determinant of $\m{B}$)
remain the same. Note that in~\eqref{solSspec3Cl30etro} the scalar
product in \cl{1}{2} has both plus/minus signs, in particular
$\bv^2=v_1^2-v_2^2-v_3^2$. Before considering other algebras it is
helpful to analyze few examples.
\subsection{Examples for $\cl{3}{0}$ and $\cl{1}{2}$}
\subsubsection*{Example~1. The case $s\ne S$.}\label{exmp1}
The square root of $\B=\e{1}-2 \e{23}$ in $\cl{3}{0}$. The
coefficients in this case are $b_1=1$ and $b_{23}=-2$, and all
remaining ones are equal to zero. Then, from \eqref {bSbICl30}
follows that  $b_I=-4$ and $b_S=3$. The expression
\eqref{solBCl30} gives $t_{1,2}=(\tfrac{1}{4},-\tfrac{1}{4})$ and
$T_{1,2}=(-\tfrac{1}{2},\tfrac{1}{2})$.  Finally, using
\eqref{solACl30} we find the real solutions for $s$ and $S$,
\begin{equation}\begin{split}\label{ex1}
  &\bigl(s_{1,2}=\mp\tfrac{1}{2}c_1,\ S_{1,2}=\pm\tfrac{1}{2}c_2\bigr)\quad \text{and}\quad
  \Bigl(s_{3,4}=\pm\tfrac{1}{2}c_2,\ S_{3,4}=\pm\tfrac{1}{2}c_1\Bigr),
\end{split}\end{equation}
where $c_1=\sqrt{-2+\sqrt{5}}$ and  $c_2=\sqrt{2+\sqrt{5}}$. Thus,
the MV is regular. Using \eqref{vV1Cl30}-\eqref{vV2Cl30} then we
have the following four sets of non-zero coefficients:
\begin{equation}\label{ex1aCl30}\begin{array}{llll}
\bigl(s_1=-\tfrac{1}{2}c_1,& S_1=\tfrac{1}{2}c_2,& v_1=-\tfrac{1}{2}c_2,& V_1=-\tfrac{1}{2}c_1\bigr),\\
\bigl(s_2=\tfrac{1}{2}c_1,& S_2=-\tfrac{1}{2}c_2,& v_1=\tfrac{1}{2}c_2,& V_1=\tfrac{1}{2}c_1\bigr),\\
\bigl(s_3=\tfrac{1}{2}c_2,& S_3=\frac{1}{2}c_1,& v_1=\tfrac{1}{2}c_2,& V_1=-\tfrac{1}{2}c_1\bigr),\\
\bigl(s_4=-\tfrac{1}{2}c_2,& S_4=-\tfrac{1}{2}c_1,&
v_1=-\tfrac{1}{2}c_1,& V_1=\tfrac{1}{2}c_2\bigr).
\end{array}\end{equation}
The remaining coefficients are equal to zero, $v_2=v_3=V_2=V_3=0$.
Finally, inserting the coefficients~\eqref{ex1aCl30} into
\eqref{mvA30} one can find  four different  roots,
\begin{equation}\begin{split}\label{ex1fa}
  &\A_{1,2} = \mp\tfrac12 c_2 \bigl(-2+\sqrt{5}+\e{1}+(-2+\sqrt{5}) \e{2 3}-\e{1 2 3}\bigr),\\
  &\A_{3,4}= \pm \tfrac12 c_1 \bigl(2+\sqrt{5}+\e{1}-(2+\sqrt{5}) \e{2 3}+\e{1 2
  3}\bigr),
\end{split}\end{equation}
squares of which  give the initial MV $\B=\e{1}-2 \e{23}$.

\subsubsection*{Example~2. The case $s=S$.\label{exmp2}}
The square root of $\B=-1+\e{3}-\e{12}+\frac12\e{123}$ in
$\cl{3}{0}$. Now $b_0=-1$, $b_{123}=\tfrac12$, $b_I=-1$,
$b_S=\tfrac34$. Then, from \eqref{solACl30} and \eqref{solBCl30}
follows real solutions for $s_i$ and $S_i$,
\begin{equation}\label{ex2}
\bigl(s_{1,2}=\pm\tfrac{1}{2},\quad S_{1,2}=\pm\tfrac{1}{2}\bigr)\quad\textrm{and}\quad \bigl(s_{3,4}=0,\quad S_{3,4}=\pm 1\bigr).
\end{equation}
Then, for case $(s_{1,2},S_{1,2})$ the equations
\eqref{vV1Cl30}-\eqref{vV2Cl30} yield
\begin{equation}\begin{array}{llll}\label{ex2a}
\bigl(s_1=-\tfrac{1}{2},&\ v_1=v_2=v_3=0,&\ V_1=V_2=0,&\ V_3=1\bigr),\\
\bigl(s_2=\tfrac{1}{2},&\ v_1=v_2=v_3=0,&\ V_1=V_2=0,&\ V_3=-1\bigr).\\
\end{array}\end{equation}
The case $(s_{3,4},S_{3,4})$ is treated exactly as in
\textit{Example}~1. The final answer consists of four roots too,
\begin{equation}\begin{split}\label{ex2fa}
  &\A_{1,2}= \pm \tfrac12\bigl(-1 +2\e{12}-\e{1 2 3}\bigr),\\
  &\A_{3,4}= \pm \tfrac12\bigl(\e{3} +  \e{12}-2\e{1 2 3}\bigr).
\end{split}\end{equation}

\subsubsection*{Example~3. The case $s=S=0$.}\label{ex3Cl30}
The square root of $\B=-1+\e{123}$, which is the center of
$\cl{3}{0}$. The coefficients $b_0=-1$, $b_{123}=1$ give
$b_I=2$, $b_S=0$. Then, from expressions \eqref{solBCl30} and
\eqref{solACl30} follows
\begin{equation}\label{ex3}\begin{split}
&\bigl(s_{1,2}=\pm c_1,\ S_{1,2}=\pm c_2\bigr)\quad
\text{and}\quad  \bigl(s_{3}=0,\ S_{3}=0\bigr),
\end{split}\end{equation}
where  $c_1=\sqrt{-1/2+1/\sqrt{2}}$ and
$c_2=\sqrt{1/2+1/\sqrt{2}}\,$.

The case $(s_{1,2}, S_{1,2})$ in \eqref{ex3} can be computed
similarly as in \textit{Example}~1. The two square roots, which
are obtained from case $\bigl(s_{1,2}=\pm c_1,\ S_{1,2}=\pm
c_2\bigr)$, therefore  are
\begin{equation}
\A_{1,2}= \pm (c_1+c_2 \e{123}).
\end{equation}

The set of two roots above should be extended by adding a set of
roots provided by the case $(s_{3}=0,\ S_{3}=0)$ in \eqref{ex3},
which is special. Indeed, some of coefficients in this case remain
unspecified and therefore may be treated as free parameters that
yield an uncountable number (continuum) of roots. The coefficients
($b_1, b_2, b_3$) and ($b_{12}, b_{13}, b_{23}$) in this case are
zeroes, however, the compatibility of
\eqref{MVM2Cl30}-\eqref{MVM4Cl30} is satisfied and the solution
set is not empty. Indeed, as seen from \eqref{solSspec3Cl30etro}
the system can  be solved for an arbitrary pair of coefficients
($v_1,v_2,v_3,V_1,V_2,V_3$), for example with ($v_1,V_1$). If
($v_1,V_1$) is inserted into~\eqref{mvA30} one gets MV with four
free parameters,
\begin{equation}
\A=f_1(v_2,v_3,V_2,V_3)\e{1}+v_2\e{2}+v_3\e{3}+f_2(v_2,v_3,V_2,V_3)\e{23}-V_2\e{13}+V_3\e{12},
\end{equation}
 where $v_1=f_1(v_2,v_3,V_2,V_3)$ and
$V_1=f_2(v_2,v_3,V_2,V_3)$ denote explicit solutions
of~\eqref{solSspec3Cl30etro},
\begin{equation}\label{exampleCl30free}
\begin{split}
&v_{1}=\mp \frac{c_1}{\sqrt{2}},\qquad
  V_{1}=\pm \frac{1}{c_1} \frac{-b_{123}+2 (v_{2} V_{2}+v_{3} V_{3})}{\sqrt{2}},\qquad \textrm{with}\quad \\[3pt]
&c_1=\Bigl(\pm\sqrt{\left(b_{0}-v_{2}^2-v_{3}^2+V_{2}^2+V_{3}^2\right)^2+
   (b_{123}-2 (v_{2} V_{2}+v_{3} V_{3}))^2}\\
&\phantom{\frac{1}{\sqrt{2}}b_{0}-v_{2}^2-v_{3}^2  (b_{123}-2
(v_{2} V_{2}+v_{3} V_{3}))x}
 +\left(b_{0}-v_{2}^2-v_{3}^2+V_{2}^2+V_{3}^2\right)\Bigr)^{\frac{1}{2}}.
\end{split}
\end{equation}
For example, by setting all free parameters to zero,
$v_2=V_2=v_3=V_3=0$, we select from a continuum two roots, which
we denote
\begin{equation}\label{exampleCl30freecontroot}
 \A_{3,4}= \pm (c_1\e{1}+c_2 \e{23}).
\end{equation}
It is important to realize, however, that the number
of roots provided by case $(s_{3}=0,\ S_{3}=0)$ in general is
infinite and the two roots
in~\eqref{exampleCl30freecontroot} represent the simplest
choice of free parameters. All roots $\A_{j}$ satisfies
$\A_{j}^2=\B=-1+\e{123}$.

If instead of $\B=-1+\e{123}$ we would have tried to find
the square root of MV that does not belong to the center, for
example have worked with $\B=\e{1}+\e{12}$, which is
directly related to polarized electromagnetic wave in $\cl{3}{0}$
\cite{Dargys2012}, we would have ended up with an empty solution
set. Indeed, in the latter case  $s_1=0$, $S_1=0$ and
$b_0=b_{123}=b_I=b_S=0$. Then, after substitution of $s\to s_1=0$
and $S\to S_1=0$ into Eqs.~\eqref{MVM2Cl30}-\eqref{MVM4Cl30} one
obtains the contradiction, $1=0$.

\subsubsection*{Example~4. The case of quaternion.}
The quaternions are isomorphic to even subalgebra $\cl{3}{0}^{+}$
with  elements $\{1,\e{12},\e{23}, \e{13}\}$, therefore, the
provided formulas allow to find quaternionic square root too.
Taking into account that quaternion imaginary units are
$\mathbf{i} =\e{12}$, $\mathbf{j}=-\e{13}$ and
$\mathbf{k}=\e{23}$,  let's compute the square root of
$\B=1+\e{12}-\e{13}+\e{23}=1+\mathbf{i}+\mathbf{j}+\mathbf{k}$. In
this example  we have  $b_0=1$, $b_{123}=0$ and $b_I=0$, $b_S=4$.
Starting from \eqref{solBCl30} and then using Eq.~\eqref{solACl30}
it is easy to find that the MV represents a regular case with four
different coefficients
\begin{equation}\begin{split}\label{ex4}
  &\bigl(s_{1,2}=0,\quad S_{1,2}=\pm 1/\sqrt{2}\bigr)\quad \text{and}\quad
  \Bigl(s_{3,4}=\pm \sqrt{3/2},\quad S_{3,4}=0\Bigr).
\end{split}\end{equation}
Using \eqref{vV1Cl30}-\eqref{vV2Cl30} and \eqref{mvA30} we can write the answer:
\begin{equation}\begin{split}\label{ex4fa}
  &\A_{1,2}= \pm (\e{1} +\e{2}+\e{3}+\e{1 2 3})/\sqrt{2}\,,\\
  &\A_{3,4}= \pm(3 +\e{12}-\e{13}+\e{23})/\sqrt{6}\equiv\pm(3 +\mathbf{i}+\mathbf{j}+\mathbf{k})/\sqrt{6}\,.
\end{split}\end{equation}
The squares  of all roots yield the initial MV. It should be
noticed that in $\A_{3,4}$  the quaternion imaginary units
have remained in even algebra only. The source of this
`strange' difference is related with the algorithm used in \ref{appendix:algorithm}, where the roots are
computed by \cl{3}{0} algebra rather than by \cl{0}{2}, i.e. to
algebra of quaternions. However, the program for \cl{0}{2} (see
resp. equation in \ref{appendix:dim2}) gives two roots only.

\subsubsection*{Example~5. The regular case of $\cl{1}{2}$ algebra.}
Using the same initial MV,  $\B=\e{1}-2 \e{23}$  as in
\textit{Example}~1, one obtains the same values for $(b_S, b_I)$
and $(s, S)$. After substitution into \eqref{vV1Cl12},
\eqref{vV2Cl12} and then into \eqref{mvA30} the square roots are
found to be
\begin{equation}\begin{split}
  &\A_{1,2}=\pm\tfrac12(c_2 (-\e{1}+\e{1 2 3})-c_1 (1+\e{2 3})),\\
  &\A_{3,4}=\pm \tfrac12(-c_1 (\e{1}+\e{1 2 3})+c_2 (-1+\e{2 3})),
\end{split}\end{equation}
where $c_1=\sqrt{-2+\sqrt{5}}$ and  $c_2=\sqrt{2+\sqrt{5}}$\,.

\section{Square roots in \cl{0}{3} algebra}\label{sec:Cl03}
The similar approach to the root problem allows to write down
explicit square root formulas for \cl{0}{3} algebra as well. Using
the same notation~\eqref{mvA30} for $\A$ and $\B$ and equating
coefficients at same basis elements in $\A^2=\B$ now we obtain the
following system of equations,\footnote{The formulas have the same
structure and  differ in signs of some constituent terms only.
Below, for easier reading and application, all formulas, including
those for mixed algebras,  are written explicitly without
introducing a large number of sign epsilons $\varepsilon_{\pm}=\pm
1$. In fact, appearance of different signs in structurally similar
expressions brings in different conditions for  real root
existence in distinct algebras.}
\begin{align}
b_0&=s^2+S^2+\bv^2+\bV^2, &b_{123}&=2(sS+\bv\cdot\bV)\label{MVM10Cl03}, \\
b_1&=2(s v_1+S V_1), &b_{23}&=-2(s V_1+S v_1)\label{MVM20Cl03}, \\
b_2&=2(s v_2+S V_2), &b_{13}&=2(s V_2+S v_2)\label{MVM30Cl03}, \\
b_3&=2(s v_3+S V_3), &b_{12}&=-2(s V_3+S v_3)\label{MVM40Cl03},
\end{align}
where now $\bv^2=-v_1^2-v_2^2-v_3^2$ and $\bv\cdot\bV=-v_1 V_1- v_2 V_2- v_3 V_3$.

\subsection{The generic case $s^2-S^2\neq 0$}
The solution of  Eqs.~\eqref{MVM20Cl03}-\eqref{MVM40Cl03} is
\begin{align}
v_1&=\frac{b_1s+b_{23}S}{2(s^2-S^2)}, & v_2&=-\frac{b_{2}s- b_{13}S}{2(s^2-S^2)},&  v_3&=\frac{b_3s+b_{12}S}{2(s^2-S^2)},\label{v123Cl03}\\
V_1&=-\frac{b_{23}s+b_{1}S}{2(s^2-S^2)},
&V_2&=\frac{b_{13}s-b_{2}S}{2(s^2-S^2)},
&V_3&=-\frac{b_{12}s+b_{3}S}{2(s^2-S^2)},\label{V123Cl03}
\end{align}
which is valid when $s^2-S^2\neq 0$, and corresponds to the
generic case.  After substitution of \eqref{v123Cl03} and
\eqref{V123Cl03} into \eqref{MVM10Cl03} one obtains two coupled
nonlinear algebraic equations for two unknowns $s$ and $S$,
\begin{equation}\label{sysA0Cl03}
\begin{split}
& b_{S}+4 s^2 (-6 S^2+b_0)+8 s S b_{123}=4 s^4+(-2 S^2+b_0)^2+b_{123}^2,\\
  & b_{I}=2 (2 (s^2+S^2)-b_0) (4 s S-b_{123}),
\end{split}\end{equation}
where  again the coordinate-free notation is introduced,
\begin{equation}\label{bSbICl03}
\begin{split}
&b_{S}= \langle \B \cliffordconjugate{\B}\rangle_0 =
b_{0}^2+b_{1}^2+b_{2}^2+b_{3}^2+ b_{12}^2+b_{13}^2+b_{23}^2+b_{123}^2,\\
&b_{I}= \langle \B
\cliffordconjugate{\B} I \rangle_0 = -2 b_{3} b_{12}+2 b_{2} b_{13}-2 b_{1} b_{23}+2 b_{0} b_{123}.
\end{split}
\end{equation}
Note change of signs as compared to \cl{3}{0} case. The
determinant $D$ in \cl{0}{3} is expressed as a difference,
$D=b_S^2-b_I^2$, which is always positive, $D>0$.

To reduce the degree of the above equations, the substitution
\begin{equation}\label{sysAS0Cl03}
  s\,S = t;\qquad  \tfrac{1}{2}(s^2 + S^2) = T,
\end{equation}
is used that transforms the system~\eqref{sysA0Cl03} into simpler
one,\footnote{To eliminate $s$ and $S$, {\it Mathematica} commands
\textbf{Eliminate[~]},
  \textbf{GroebnerBasis[~]} have been used. They allow to rewrite the
initial  Eqs~\eqref{sysAS0Cl03}  in a number of equivalent forms.}
\begin{equation}\label{sysAR0Cl03}
 b_{S}=(4 T-b_0)^2+(4 t-b_{123})^2,\qquad
  b_{I}=2 (4 T-b_0) (4 t-b_{123}).
\end{equation}
The solution of~\eqref{sysAR0Cl03} when $(b_{S}\pm\sqrt{D})
> 0$ is
\begin{align}\label{solB0Cl03}
&\begin{cases}
 \Bigl(t_{1,2}=\frac{1}{4} \bigl(b_{123} \pm \frac{1}{\sqrt{2}}\sqrt{b_{S}-\sqrt{D}}\bigr),\quad
  T_{1,2}=\frac{1}{4}\bigl(b_{0}\pm\frac{ b_{I}}{\sqrt{2}\sqrt{b_{S}-\sqrt{D}}}\bigr)\Bigr),\\
  \hphantom{t_{1,2}=\frac{1}{4} \Bigl(b_{123} \pm \frac{1}{\sqrt{2}}\sqrt{-b_{S}+\sqrt{D}}\Bigr),\
  T_{1,2}=\frac{1}{4}b_{123}\quad\;}
  \quad \textrm{if}\ b_{S}-\sqrt{D}>0,
  \\[2pt]
  \bigl(t_{1,2}=\frac{1}{4}b_{123},\quad   T_{1,2}=\frac{1}{4} (\pm\sqrt{b_{S}}+b_{0})\bigr),
    \ \textrm{if}\quad b_{S}-\sqrt{D}=0\ \textrm{and}\ b_{S}>0.
\end{cases}
\end{align}
The $\pm$ signs in the above formulas are mutually related,
thus, there are only two possibilities that correspond to either
plus or minus signs inside $t_i$ and $T_i$ formulas. The remaining two solutions of
\eqref{sysAR0Cl03}, which were obtained from \eqref{solB0Cl03}
after replacement $-\sqrt{D}\to +\sqrt{D}$, yield a complex valued
expression for $T\pm\sqrt{T^2- t^2}$ (see Eq.~\eqref{solA0Cl03}
below) therefore they were dismissed in advance.

Once the equations in~\eqref{solB0Cl03} are computed they can be
substituted back into solutions of~\eqref{sysAS0Cl03},
\begin{equation}\label{solA0Cl03}
\begin{cases}
  \Bigl(s_{1,2,3,4} =\pm\sqrt{T\pm\sqrt{T^2- t^2}},
\ S_{1,2,3,4}
  =\frac{\pm t}{\sqrt{T\pm\sqrt{T^2- t^2}}}\Bigr) \textrm{if}\ T\ge 0,\ t\neq 0,\\[2pt]
 (s_{1,2}= S_{3,4} =\pm\sqrt{2T},\qquad S_{1,2}= s_{3,4} = 0) \qquad\quad\quad \textrm{if}\quad T\ge 0,\ t= 0.
\end{cases}\end{equation}
In the obtained equations the same signs must be chosen  in the
same index positions in $s_{1,2,3,4}$ and $S_{1,2,3,4}$ (four
possibilities).

Thus, starting  from pairs $(t_1,T_1)$ and $(t_2,T_2)$ in
Eq.~\eqref{solB0Cl03} and then going to \eqref{solA0Cl03},
and finally to formulas \eqref{v123Cl03}, \eqref{V123Cl03} and
\eqref{mvA30} one obtains explicit real solutions that completely
determine the square root of equation $\B=\A^2$ (with
$\A=s+\bv+(S+\bV)I$) of the generic MV $\B$ of real $\cl{0}{3}$
algebra in radicals. It appears that, at most, only four real
solutions\footnote{Because the condition $T\ge 0$ in
\eqref{solA0Cl03} selects only single sign from
\eqref{solB0Cl03}.} are possible in this algebra too, since other
choices of signs in \eqref{solB0Cl03} and \eqref{solA0Cl03} yield
negative expressions inside square roots.

\subsection{The special case $s^2 - S^2=0$}

There are three subcases:  1)~$s = S\neq 0$,  2)~$s =
-S\neq 0$ and 3) $s=S=0$.
\subsubsection{The subcase $s=S\neq 0$\label{s2=S2}}
Here the system of  Eqs.~\eqref{MVM20Cl03}--\eqref{MVM40Cl03} has
a special solution,
\begin{equation}\begin{split}\label{vVSpecialCl03}
v_{1}= \frac{b_{1}}{2 s}-V_{1},\quad v_{2}= \frac{b_{2}}{2 s}-V_{2},\quad v_{3}= \frac{b_{3}}{2 s}-V_{3},
\end{split}\end{equation}
 if and only if the MV $\B$
coefficients satisfy: $b_1 =-b_{23}$, $b_2 = b_{13}$,  $b_3= -
b_{12}$. In \eqref{vVSpecialCl03}, $v_i$ is expressed in
terms of $V_i$. Appearance of $s$ in the denominators implies
that the case $s=S=0$ must be investigated separately. After
substituting the solution \eqref{vVSpecialCl03} into
\eqref{MVM10Cl03} and taking into account the mentioned conditions
($b_1 =- b_{23}, b_2 = b_{13}, b_3= - b_{12}$) one gets two
equations,
\begin{equation}\begin{split}\label{sysAspecCl03}
  &-\frac{b_{1}^2+b_{2}^2+b_{3}^2}{4 s^2}+\frac{b_{1} V_{1}+b_{2}
  V_{2}+b_{3} V_{3}}{s}-b_{0}+2 s^2+2 (\bV\cdot \bV)=0,\\
&-\frac{b_{1} V_{1}+b_{2}  V_{2}+b_{3} V_{3}}{s}-b_{123}+2 s^2-2 (\bV\cdot \bV)=0,
\end{split}\end{equation}
that should be kept mutually compatible. To this end we subtract
and add the above equations to get
\begin{equation}\begin{split}\label{sysAspecSubsAddCl03}
  &-\frac{4 s^2 \left(b_{0}-b_{123}-4 (\bV\cdot \bV)\right)-8 s (b_{1} V_{1}+b_{2} V_{2}+b_{3} V_{3})+b_{1}^2+b_{2}^2+b_{3}^2}{4 s^2}=0,\\
  &  -\frac{4 s^2 \left(b_{0}+b_{123}-4 s^2\right)+b_{1}^2+b_{2}^2+b_{3}^2}{4 s^2}=0.
\end{split}\end{equation}
Then, making use of  expanded form of  $b_S=\langle \B
  \cliffordconjugate{\B}\rangle_0=b_0^2 + 2 (b_1^2 + b_2^2 +
b_3^2) + b_{123}^2$, where the conditions $b_1 =- b_{23}, b_2 =
b_{13},  b_3= - b_{12}$ have been taken into account, one can
express the sum $b_{1}^2+b_{2}^2+b_{3}^2$ from the second equation
in~\eqref{sysAspecSubsAddCl03}, $b_{1}^2+b_{2}^2+b_{3}^2 =\frac12
(b_S-b_{0}^2-b_{123}^2)$, and substitute the latter into
the first of equations. The result is the quadratic equations for
$V_i$'s. After solving, for example, with respect to $V_1$, one
can express $V_1$ in terms of, now, arbitrary free parameters
$V_2$ and $V_3$,
\begin{equation}\begin{split}\label{AspecCl03}
  V_{1}=& \frac{\sqrt{2}}{8 s}\Bigl(\sqrt{2} b_{1}\pm \bigl(-8 s^2 \left(b_{0}-b_{123}+4 \left(V_{2}^2+V_{3}^2\right)\right)+\\
  &\phantom{\bigl(-8 s^2 }
  16 s (b_{2} V_{2}+b_{3} V_{3})+b_{0}^2+2 b_{1}^2+b_{123}^2-b_{S}\bigr)^{1/2}\Bigr),
\end{split}\end{equation}
that warrants compatibility of the system
\eqref{sysAspecCl03}. Thus, further analysis may be restricted to
the  simplest single equation,  the second equation in
\eqref{sysAspecSubsAddCl03} that after introduction of shortcut
$b_S$ can be cast to form,
  \begin{equation}\begin{split}\label{sysAspec2Cl03}
b_{S}=-8 b_{0} s^2-8 b_{123} s^2+b_{0}^2+b_{123}^2+32 s^4 .
 \end{split}\end{equation}
Solution of \eqref{sysAspec2Cl03} with respect to $s$ can be
expressed in radicals,
\begin{equation}\begin{split}\label{solSspec1Cl03}
s_{1,2} =& \pm\tfrac{1}{2 \sqrt{2}}\sqrt{\sqrt{2
b_{S}-(b_{0}-b_{123})^2}+b_{0}+b_{123}}\,,
\end{split}\end{equation}
where all expressions inside square roots are assumed to be
positive. The expressions \eqref{solSspec1Cl03}, \eqref{AspecCl03}
and \eqref{vVSpecialCl03} after substitution into \eqref{mvA30}
yield the final answer for this special case under conditions for
MV $\B$ coefficients: $b_1 =- b_{23},  b_2 = b_{13},  b_3= -
b_{12}$ that  in an abridged version can be reduce to
$b_S-(b_0-b_{123})^2=b_I$. In conclusion, the solution set
contains two free parameters, $V_2$ and $V_3$, and therefore
represents continuum of roots on a two dimensional manifold in the
parameter space.

\subsubsection{The subcase $s=-S\neq 0$}
Performing exactly the same analysis as in subcase~\ref{s2=S2},
one obtains  the conditions for existence of solution:  $b_1 =
b_{23}$, $b_2 = -b_{13}$ and $b_3= b_{12}$, or in short
$-b_S+(b_0+b_{123})^2=b_I$. Similarly, expressing $v_i$ in terms
of $V_i$ one gets
\begin{equation}\begin{split}\label{vVSpecial2Cl03}
v_{1}= \frac{b_{1}}{2 s}+V_{1},\quad v_{2}= \frac{b_{2}}{2 s}+V_{2},\quad v_{3}= \frac{b_{3}}{2 s}+V_{3}.
\end{split}\end{equation}
If  $V_1$ is expressed in terms of $V_2$ and $V_3$,
\begin{equation}\begin{split}\label{Aspec2Cl03}
  V_{1}=& \frac{\sqrt{2}}{8 s}\Bigl(-\sqrt{2} b_{1}\pm \big(-8 s^2 \big(b_{0}+b_{123}+4(V_{2}^2+V_{3}^2)\big)\\
  &\phantom{\big(-8 s^2 }
  -16 s (b_{2} V_{2}+b_{3} V_{3})+b_{0}^2+2 b_{1}^2+b_{123}^2-b_{S}\big)^{1/2}\Bigr),
\end{split}\end{equation}
we find two real solutions for $s$,
\begin{equation}\begin{split}\label{solSspec1NegCl03}
s_{1,2} =& \pm\tfrac{1}{2 \sqrt{2}}\sqrt{\sqrt{2
b_{S}-(b_{0}+b_{123})^2}+b_{0}-b_{123}}\,,
\end{split}\end{equation}
After substitution into \eqref{mvA30} the above expressions again
yield the final MV provided the conditions $b_1 = b_{23},  b_2 =
-b_{13}, b_3= b_{12}$ are satisfied.

\subsubsection{The subcase $s = S = 0$}
The analysis of this special subcase is very similar to that in
$\cl{3}{0}$. The Eqs.~\eqref{MVM20Cl03}-\eqref{MVM40Cl03} satisfy
compatibility condition if   vector $(b_1, b_2, b_3)$ and bivector
$(b_{12}, b_{13}, b_{23})$ coefficients  are equated to zero. Then
the Eqs.~\eqref{MVM10Cl03} assume the following form
\begin{equation}\label{solSspec3Cl03}
b_0=\bv^2+\bV^2,\quad b_{123}=2(\bv\cdot\bV)
\end{equation}
from which follows that  four parameters remain unspecified. For
example, if Eq.~\eqref{solSspec3Cl03} is solved with respect to
pair $(v_1,V_1)$, one gets
\begin{equation}\begin{split}
&v_{1}=\mp \frac{c_1}{\sqrt{2}},\qquad
  V_{1}=\pm \frac{1}{c_1} \frac{b_{123}+2 (v_{2} V_{2}+v_{3} V_{3})}{\sqrt{2}},\qquad \textrm{where}\quad \\[3pt]
&c_1=\Bigl(\pm\sqrt{\left(b_{0}+v_{2}^2+v_{3}^2+V_{2}^2+V_{3}^2\right)^2-
   (b_{123}+2 (v_{2} V_{2}+v_{3} V_{3}))^2}\\
&\phantom{\frac{1}{\sqrt{2}}b_{0}-v_{2}^2-v_{3}^2  (b_{123}-2
(v_{2} V_{2}+v_{3} V_{3}))x}
-b_{0}-v_{2}^2-v_{3}^2-V_{2}^2-V_{3}^2\Bigr)^{\frac{1}{2}}.
\end{split}
\end{equation}
The pairs  $(v_2,V_2)$ and $(v_3,V_3)$ may be interpreted as free
parameters that generate a continuum of roots in a four parameter
space. The geometric interpretation of
Eqs.~\eqref{solSspec3Cl03} is similar to those
in~\eqref{solSspec3Cl30etro} for \cl{3}{0}.

\subsection{Examples for $\cl{0}{3}$}
\subsubsection*{Example~6. The regular case.}

As in \textit{Example}~1, let the initial MV be $\B=\e{1}-2
\e{23}$, the coefficients of which  are $b_1=1$, $b_{12}=-2$. The
shortcuts $b_I$ and $b_S$ in~\eqref{bSbICl03} have the values
$b_I=4$ and $b_S=5$. The formulas in~\eqref{solB0Cl03} give
$(T_{1},t_{1})= (\tfrac{1}{2},\tfrac{1}{4})$ and $(T_{2},t_{2})=
(-\tfrac{1}{2},-\tfrac{1}{4})$. Since $T_1$ is positive, the pair
$(T_{1},t_{1})$ is used in the following. Then, from the system
\eqref{solA0Cl03} we find four values of $(s_i,S_i)$ and then from
\eqref{v123Cl03}, \eqref{V123Cl03} the coefficients $v_i$ and
$V_i$,
\begin{equation}\label{ex1aCl03}
\begin{array}{lll}
\big(s_1=V_1=\tfrac{1}{4}d_3,&S_1=v_1=-\tfrac{1}{2}d_2,&v_2=v_3=V_2=V_3=0\big),\\
\big(s_2=V_1=\tfrac{1}{2}d_1,&S_2= v_1=\tfrac{1}{2}d_2,&v_2=v_3=V_2=V_3=0\big),\\
\big(s_3=V_1=-\tfrac{1}{2}  d_2,&S_3=-\tfrac{1}{2 d_2},\ v_1=\tfrac{1}{4}d_3,&v_2=v_3=V_2=V_3=0\big),\\
\big(s_4= V_1=\tfrac{1}{2}d_2,&S_4=\tfrac{1}{2d_2},\ v_1=\tfrac{1}{2}d_1,& v_2=v_3=V_2=V_3=0\big).\\
\end{array}
\end{equation}
where $d_1=\sqrt{2-\sqrt{3}}$,  $d_2=\sqrt{2+\sqrt{3}}$ and
$d_3=\sqrt{2}-\sqrt{6}$. Finally, in the same way as in
\textit{Example}~1 we find four different square roots,
\begin{equation}\begin{split}\label{ex1AnsCl03}
 &\A_{1}=\tfrac{1}{2}\big(d_1+d_2 \e{1}-d_1 \e{23}+ d_2\e{1 2 3}\big),\\
 &\A_{2}=\tfrac{1}{2}\big(\tfrac{1}{2}d_3-d_2\e{1}-\tfrac{1}{2}d_3 \e{23} -d_2\e{1 2 3}\big).\\
 &\A_{3}=\tfrac{1}{2}\big(d_2+ d_1\e{1}-d_2\e{2 3}+\frac{\e{123}}{d_2}\big),\\
 &\A_{4}=\tfrac{1}{2}\big(-d_2+ \tfrac{1}{2}d_3\e{1}+d_2\e{2 3} -\frac{\e{1 2 3}}{d_2}\big),\\
\end{split}\end{equation}
Noting that $d_3=-2d_1$ and $d_2^{-1}=d_1$ the roots may
rewritten in the standard form
\begin{equation}\begin{split}\label{ex1AnsCl03B}
 &\A_{1,2}=\pm\tfrac{1}{2}\big(d_1+d_2 \e{1}-d_1 \e{23}+ d_2\e{1 2 3}\big),\\
 &\A_{3,4}=\pm\tfrac{1}{2}\big(d_2+ d_1\e{1}-d_2\e{2 3}+d_1\e{123}\big).\\
\end{split}\end{equation}

\subsubsection*{Example~7. The case $s= S\neq 0$.}
The square root of  $\B=-\e{3}+\e{12}+4\e{123}$. The shortcuts in
$(b_I,b_S)$ have values $b_I=2$ and $b_S=18$, and afterwards the
expression \eqref{solB0Cl03} gives $(T_{1},t_{1})=
(\tfrac{c_1}{4},\tfrac{c_1}{4})$, where $c_1=(2+\sqrt{5})$. The
negative $T$ solution has been omitted. All this gives
$(s_1,S_1)=(-\tfrac{\sqrt{c_1}}{2},-\tfrac{\sqrt{c_1}}{2})$ and
$(s_2,S_2)=(\tfrac{\sqrt{c_1}}{2},\tfrac{\sqrt{c_1}}{2})$. The
coefficients satisfy the relations $b_1 =- b_{23}$,  $b_2 =
b_{13}$,  $b_3= -b_{12}$, therefore, a special solution
consisting of four MVs  exists:
\begin{equation}\label{ex2AnsCl03}
\begin{split}
  &\A_1=-\frac{1}{2} (\e{1}+\e{2 3}) \sqrt{-4 V_2^2+4 (c_1-V_3) V_3+c_3}-\e{1 2} V_3+(\e{1 3}-\e{2}) V_2\\
  &\hphantom{-\frac{1}{2} (\e{1}+\e{2 3})\sqrt{-4 V_2^2+4 (c_1-V_3) V_3+c_3}}  +\e{3} (c_1-V_3)-\frac{1}{2} c_2 (\e{1 2 3}+1),\\
  &\A_2= \frac{1}{2} (\e{1}+\e{2 3}) \sqrt{-4 V_2^2+4
(c_1-V_3) V_3+c_3}-\e{1 2} V_3+(\e{1 3}-\e{2}) V_2\\
  &\hphantom{-\frac{1}{2} (\e{1}+\e{2 3})\sqrt{-4 V_2^2+4 (c_1-V_3) V_3+c_3}} +\e{3} (c_1-V_3)-\frac{1}{2} c_2 (\e{1 2 3}+1),\\
  &\A_3=  \frac{1}{2} (-(\e{1}+\e{2 3}) \sqrt{-4 V_2^2-4 V_3 (V_3+c_1)+c_3}-2 \e{1 2} V_3\\
  &\hphantom{-\frac{1}{2} (\e{1}+\e{2 3})\sqrt{-4 V_2^2}\ } +2 (\e{1 3}-\e{2})
V_2-2 \e{3} (V_3+c_1)+c_2 (\e{1 2 3}+1)),\\
  &\A_4=
  \frac{1}{2} ((\e{1}+\e{2 3}) \sqrt{-4 V_2^2-4 V_3 (V_3+c_1)+c_3}-2 \e{1 2} V_3\\
  &\hphantom{-\frac{1}{2} (\e{1}+\e{2 3})\sqrt{-4 V_2^2}\ }  +2 (\e{1 3}-\e{2}) V_2-2 \e{3} (V_3+c_1)+c_2 (\e{1 2 3}+1)),
\end{split}
\end{equation}
where $c_1=\sqrt{\sqrt{5}-2}$,  $c_2=\sqrt{\sqrt{5}+2}$,
$c_3=-\sqrt{5}+6$. Assuming concrete values of parameters $V_2$
and $V_3$ one can check that the root formulas give real MVs.

It should be noted however that symbolical expressions do not
guarantee that we will always be able to find {\it real}
parameters $V_2$ and $V_3$, what would ensure real square roots.
For example, if instead of the above MV
$\B=-\e{3}+\e{12}+4\e{123}$ we would try to find square root of MV
$\B=-\e{3}+\e{12}$ in \cl{0}{3} (the MV was used earlier in
the \textit{Example~2}) we would find $s_1=\tfrac{1}{2}$ and
$s_2=-\tfrac{1}{2}$. The first value then yields
$v_1=-V_1=\tfrac{1}{2} \sqrt{-4 V_2^2-(1+2 V_3)^2}$, $v_2=-V_2$,
$v_3=-1-V_3$ and the second one yields $V_1=-v_1=\tfrac{1}{2}
\sqrt{-4 V_2^2-(1-2 V_3)^2}$, $v_2=-V_2$, and $v_3=1-V_3$.
 Taking the square of symbolical expressions one can easily
check that formally we indeed obtain the MV $\m{B}=-\e{3}+\e{12}$.
It is obvious, however, that in both cases {($s_1=\tfrac{1}{2}$
and $s_2=-\tfrac{1}{2}$) the expression under square root can be
made non-negative (i.e. only zero in this case) for a single choice of
parameters. In particular, in the case $s_1=\tfrac{1}{2}$,  the requirement $-4 V_2^2-(1+2 V_3)^2\ge 0$ yields $V_2=0,
V_3=-1/2$. Alternatively, in the case $s_2=-\tfrac{1}{2}$ from
equation $-4 V_2^2-(1-2 V_3)\ge 0$ follows $V_2=0,
V_3=1/2$. The both cases yield an isolated root $\pm\frac{1}{2}(1-\e{3}+\e{12}+\e{123})$. Therefore, in this
algebra in fact there exist only isolated real square root of
$\B=-\e{3}+\e{12}$.

\section{Square roots in \cl{2}{1} algebra}\label{sec:Cl21}

\subsection{The generic case $s^2 -S^2 \neq 0$}
The system of nonlinear equations is
\begin{align}
b_0&=s^2+S^2+\bv^2+\bV^2, &b_{123}&=2(sS+\bv\cdot\bV)\label{MVM10Cl21}, \\
b_1&=2(s v_1+S V_1), &b_{23}&=2(s V_1+S v_1)\label{MVM20Cl21}, \\
b_2&=2(s v_2+S V_2), &b_{13}&=-2(s V_2+S v_2)\label{MVM30Cl21}, \\
b_3&=2(s v_3+S V_3), &b_{12}&=-2(s V_3+S v_3)\label{MVM40Cl21},
\end{align}
where now $\bv^2=v_1^2+v_2^2-v_3^2$ and $\bv\cdot\bV=v_1 V_1 + v_2
V_2- v_3 V_3$. When $s^2-S^2\neq 0$ the solutions of
system.~\eqref{MVM20Cl21}-\eqref{MVM40Cl21} are
\begin{align}
v_1&=\frac{b_1s-b_{23}S}{2(s^2-S^2)}, & v_2&=\frac{b_{2}s+ b_{13}S}{2(s^2-S^2)},&  v_3&=\frac{b_3s+b_{12}S}{2(s^2-S^2)},\label{v123Cl21}\\
V_1&=\frac{b_{23}s-b_{1}S}{2(s^2-S^2)},
&V_2&=-\frac{b_{13}s+b_{2}S}{2(s^2-S^2)},
&V_3&=-\frac{b_{12}s+b_{3}S}{2(s^2-S^2)}.\label{V123Cl21}
\end{align}
Insertion  of $v_i$ and $V_i$ into \eqref{MVM10Cl21} gives two
coupled equations for unknowns $s, S$
\begin{equation}\label{sysA0Cl21}
\begin{split}
& b_{S}+4 s^2 (-6 S^2+b_0)+8 s S b_{123}=4 s^4+(-2 S^2+b_0)^2+b_{123}^2,\\
  & b_{I}=2 (2 (s^2+S^2)-b_0) (4 s S-b_{123}),
\end{split}\end{equation}
  where $b_{S}$ and $b_{I}$ are functions of coefficients in $\B$,
\begin{equation}\label{bSbICl03A}
\begin{split}
&b_{S}= \langle \B \cliffordconjugate{\B}\rangle_0 =
b_{0}^2-b_{1}^2-b_{2}^2+b_{3}^2+ b_{12}^2-b_{13}^2-b_{23}^2+b_{123}^2,\\
&b_{I}= \langle \B
\cliffordconjugate{\B} I \rangle_0 = -2 b_{3} b_{12}+2 b_{2} b_{13}-2 b_{1} b_{23}+2 b_{0} b_{123}.
\end{split}
\end{equation}
Because the Eqs.~\eqref{sysA0Cl21} and \eqref{bSbICl03} have the same shape (the concrete equations
for $b_{S}$ and $b_{I}$, of course, are different) we can make use
of \eqref{sysAS0Cl03} with the purpose to lower the order
of the system. However, there arises an important
difference: the determinant, $D=b_S^2-b_I^2$, in \cl{2}{1} is not
always positive. It may happen that for some $\B$ the MV
determinant may become negative, $D<0$. In such a case the
solution set becomes empty. The other particularity is that in the
solution~\eqref{solB0Cl03} instead of single sign ($-\sqrt{D}$) we
have to take into account the both signs, i.e., $\pm\sqrt{D}$,
what doubles the number of possible solutions in the case $D>0$,
\begin{align}\label{solB0Cl21}
&\begin{cases}
 \Bigl(t_{1,2,3,4}=\frac{1}{4} \Bigl(b_{123} \pm
\frac{1}{\sqrt{2}}\sqrt{b_{S}\pm\sqrt{D}}\Bigr),\
T_{1,2,3,4}=\frac{1}{4}\Bigl(\frac{\pm
b_{I}}{\sqrt{2}\sqrt{b_{S}\pm\sqrt{D}}}+ b_{0}\Bigr)\Bigr),\\
  \hphantom{t_{1,2}=\frac{1}{4} \Bigl(b_{123} \pm \frac{1}{\sqrt{2}}\sqrt{-b_{S}+\sqrt{D}}\Bigr),\
  T_{1,2}=\frac{1}{4}b_{123}\ \;}
  \ \textrm{if}\quad b_{S}\pm\sqrt{D}>0,
  \\[2pt]
  \bigl(t_{1,2}=\tfrac{1}{4}b_{123},\ T_{1,2}=\tfrac{1}{4}
(\pm\sqrt{b_{S}}+b_{0})\bigr),
    \quad \textrm{if}\quad b_{S}\pm\sqrt{D}=0\ \textrm{and}\
    b_{S}>0.
\end{cases}
\end{align}
Here again the sign of $t_i$ must be taken in all possible
combinations, and the sign of $T$ must follow the same upper-lower
sign position as in  $t_i$. The condition $b_{S}\pm\sqrt{D} = 0$
implies that $b_{I}=0$. Since we already have four sign
combinations in the solution for $s,S$ (as in \eqref{solA0Cl03}),
we end up with 16~different square roots of MV in a generic case
of \cl{2}{1}.

\subsection{The special case $s^2 - S^2 =0$}

The analysis again closely follows \cl{0}{3} case, except that now
different signs appear in expressions.
\subsubsection{The subcase $s=S\neq 0$}
Now the coefficients satisfy the conditions $b_1 =b_{23}$, $b_2 =
-b_{13}$,  $b_3= - b_{12}$ which  allow to eliminate the
singularity at $s=S$. As a result the system of
Eqs.~\eqref{MVM20Cl21}-\eqref{MVM40Cl21} has a special solution,
\begin{equation}\begin{split}\label{vVSpecialCl21}
v_{1}= \frac{b_{1}}{2 s}-V_{1},\quad v_{2}= \frac{b_{2}}{2 s}-V_{2},\quad v_{3}= \frac{b_{3}}{2 s}-V_{3},
\end{split}\end{equation}
which coincides with the same solution for \cl{0}{3} (see
Eq.~\eqref{vVSpecialCl03}). Thus, after similar calculations one
finds that Eq.~\eqref{AspecCl03} becomes
\begin{equation}\begin{split}\label{AspecCl21}
  V_{1}=& \frac{\sqrt{2}}{8 s}\Bigl(\sqrt{2} b_{1}\pm \bigl(8 s^2 \left(b_{0}-b_{123}+4 (-V_{2}^2+V_{3}^2)\right)+16 s (b_{2} V_{2}-b_{3} V_{3})\\
  &\phantom{\bigl(-8 s^2 \left(b_{0}-b_{123}+4 (V_{2}^2+V_{3}^2)\right)}
  -b_{0}^2+2 b_{1}^2-b_{123}^2+b_{S}\bigr)^{1/2}\Bigr).
\end{split}\end{equation}
The coefficients $s_{1}$ and   $s_{2}$ are similar to
\eqref{solSspec1Cl03}, except that now we have to take into
account all sign combinations in inner square root,
\begin{equation}\begin{split}\label{solSspec1Cl21}
s_{1,2} =& \pm\tfrac{1}{2\sqrt{2}}\sqrt{\pm\sqrt{2
b_{S}-(b_{0}-b_{123})^2}+b_{0}+b_{123}}\,.
\end{split}\end{equation}
The above listed formulas  solve square root problem in the case
$s=S\neq 0$.

\subsubsection{The subcase $s=-S\neq 0$}
The only formulas which differ from \cl{0}{3} algebra  are
connected with the coefficient compatibility condition: $b_1 =
-b_{23}$, $b_2 = b_{13}$, $ b_3= b_{12}$. Now, the coefficients
must be replaced by
\begin{equation}\begin{split}\label{Aspec2Cl21}
  V_{1}=& \frac{\sqrt{2}}{8 s}\Bigl(-\sqrt{2} b_{1}\pm \bigl(8 s^2 \left(b_{0}+b_{123}+4(-V_{2}^2+V_{3}^2)\right)\\
  &\phantom{\bigl(-8 s^2}-16 s (b_{2} V_{2}-b_{3} V_{3})-b_{0}^2+2 b_{1}^2-b_{123}^2+b_{S}\bigr)^{1/2}\Bigr),
\end{split}\end{equation}
\begin{equation}\begin{split}\label{solSspec1NegCl21}
s_{1,2} =& \pm\tfrac{1}{2\sqrt{2}}\sqrt{\pm\sqrt{2
b_{S}-(b_{0}+b_{123})^2}+b_{0}-b_{123}}\,.
\end{split}\end{equation}
The remaining formulas which are needed for final answer exactly
match the formulas in the corresponding subcase of \cl{0}{3}
algebra.

\subsubsection{The subcase $s=S= 0$}
The only distinct formulas from \cl{0}{3} are listed below,
\begin{equation}\begin{split}
&v_{1}=\pm \frac{c_1}{\sqrt{2}},\qquad
  V_{1}=\pm \frac{1}{c_1} \frac{b_{123}+2 (-v_{2} V_{2}+v_{3} V_{3})}{\sqrt{2}},\qquad \textrm{where}\quad \\[3pt]
&c_1=\Bigl(\pm\sqrt{\left(b_{0}-v_{2}^2+v_{3}^2-V_{2}^2+V_{3}^2\right)^2-
   (b_{123}+2 (-v_{2} V_{2}+v_{3} V_{3}))^2}\\
&\phantom{\frac{1}{\sqrt{2}}b_{0}-v_{2}^2-v_{3}^2  (b_{123}-2
(v_{2} V_{2}+v_{3} V_{3}))}
+b_{0}-v_{2}^2+v_{3}^2-V_{2}^2+V_{3}^2\Bigr)^{\frac{1}{2}} .
\end{split}\end{equation}
This ends the investigation of the square root formulas for all
real 3D CAs.

\subsection{Examples for $\cl{2}{1}$}
\subsubsection*{Example~8. The regular case.}
First,  we shall show  that  MV  $\B=\e{1}-2 \e{23}$  has no real
square roots. Indeed, we have  $b_S=-5$,  $b_I=4$ and
$D=b_S^2-b_I^2=(3)^2$. As a result  the expression under square
root in \eqref{solB0Cl21}, namely $b_S\pm\sqrt{D}=-5\pm3$, is
always negative and therefore there are no real-valued solutions.

Next, we shall calculate the roots of $\B=2+\e{1}+\e{13}$. The
values of $b_I$ and $b_S$ are $0$ and $2$, respectively. The
determinant of the MV is positive, $D=4>0$. The Eqs.~\eqref{solB0Cl21} give four real
values for pairs:
 $(T_1,t_1)=\bigl(\frac{1}{4} (2-\sqrt{2}) , 0\bigr)$, $(T_2,t_2)=\bigl(\frac{1}{4} (2+\sqrt{2}) , 0\bigr)$, $(T_3,t_3)=\bigl(\frac{1}{2} , -\frac{1}{2 \sqrt{2}}\bigr)$ and $(T_4,t_4)=  \bigl(\frac{1}{2} , \frac{1}{2 \sqrt{2}}\bigr)$.
After insertion into \eqref{solA0Cl03},  16~pairs of scalars
$(s_i,S_i)$ are found:
\[ \begin{array}{lll}
 (s_1= 0,S_1=-\frac{c_2}{\sqrt{2}}),
  &\mspace{-15mu}(s_2= 0,S_2=\frac{c_2}{\sqrt{2}}),
  &\mspace{-15mu}(s_3= -\frac{c_2}{\sqrt{2}},S_3= 0),\\
  (s_4= \frac{c_2}{\sqrt{2}},S_4= 0),
  &\mspace{-15mu}(s_5= 0,S_5= -\frac{c_1}{\sqrt{2}}),
  &\mspace{-15mu}(s_6= 0,S_6= \frac{c_1}{\sqrt{2}}),\\
  (s_7= -\frac{c_1}{\sqrt{2}},S_7= 0),
  &\mspace{-15mu}(s_8= \frac{c_1}{\sqrt{2}},S_8= 0),
  &\mspace{-15mu}(s_9= -\frac{c_2}{2},S_9= \frac{c_1}{2}),\\
  (s_{10}= \frac{c_2}{2},S_{10}= -\frac{c_1}{2}),
  &\mspace{-15mu}(s_{11}= -\frac{c_1}{2},S_{11}= \frac{1}{\sqrt{2} c_1}),
  &\mspace{-15mu}(s_{12}= \frac{c_1}{2},S_{12}= -\frac{1}{\sqrt{2} c_1}),\\
  (s_{13}= -\frac{c_2}{2},S_{13}= -\frac{c_1}{2}),
  &\mspace{-15mu}(s_{14}= \frac{c_2}{2},S_{14}= \frac{c_1}{2}),
  &\mspace{-15mu}(s_{15}= -\frac{c_1}{2},S_{15}= -\frac{1}{\sqrt{2} c_1}),\\
  (s_{16}= \frac{c_1}{2},S_{16}=\frac{1}{\sqrt{2} c_1}), &
  \end{array}\]
where $c_1=\sqrt{2+\sqrt{2}}$ and $c_2=\sqrt{2-\sqrt{2}}$. After
substitution of $(s_i,S_i)$ into Eqs.~\eqref{v123Cl21} and then
into Eq.~\eqref{mvA30} we obtain 16  roots
$\A_{i,j}=\pm\sqrt{2+\e{1}+\e{13}}\,$:
\begin{equation}
\begin{split}
  &\A_{1,2}=\pm\tfrac{1}{2}\big(c_1 \e{2}-c_1 \e{2 3}-\sqrt{2}c_2\e{123}\big),\\
 &\A_{3,4}=\pm\tfrac{1}{\sqrt{2}}\big(-c_1^{-1}\e{2}+c_1^{-1}\e{2 3}+c_1 \e{1 2 3}\big),\\
&\A_{5,6}=\pm\tfrac{1}{2}\big(\sqrt{2}c_2+c_1\e{1}+c_1\e{13}\big),\\
&\A_{7,8}=\pm\tfrac{1}{\sqrt{2}}\big(c_1+ c_1^{-1}\e{1}+c_1^{-1}\e{1 3}\big),\\
&\A_{9,10}=\pm\tfrac{1}{2\sqrt{2}}\big(\sqrt{2}c_2-c_2\e{1}-c_1\e{2}-c_2\e{13}+c_1\e{2 3}+\sqrt{2}c_1 \e{1 2 3}\big),\\
  &A_{11,12}=\pm\tfrac{1}{2\sqrt{2}}\big(\sqrt{2}c_ 1+c_1 \e{1}-c_ 2 \e{2}+c_1 \e{1 3}+c_2 \e{2 3}-2c_1^{-1}\e{1 2 3}\big),\\
&\A_{13,14}=\pm\tfrac{1}{2\sqrt{2}}\big(\sqrt{2}c_1+c_1 \e{1}+c_2 \e{2}+c_1 \e{1 3}-c_2 \e{2 3}+2c_1^{-1}\e{1 2 3}\big),\\
&\A_{15,16}=\pm\tfrac{1}{2\sqrt{2}}\big(-\sqrt{2}c_2+c_2 \e{1}-c_1
\e{2}+c_2 \e{13}+c_1 \e{2 3}+\sqrt{2}c_1 \e{1 2 3}\big) .
\end{split}\end{equation}

In the end it is worth to note, that the necessary (but not
sufficient) condition for a square root of MV $\m{B}$ to exist in
real Clifford algebras $\cl{p}{q}$ requires the positivity of the
multivector determinant  $\det
(\m{B})$~\cite{Acus2018,Acus2022}. Indeed, if the MV $\m{A}$
exists and $\m{A}\m{A}=\m{B}$, then  the determinant of both sides
gives $\det (\m{A}) \det (\m{A})=\det (\m{B})$, where we have used
the multiplicative property of the determinant~\cite{Lundholm06}.
Since the determinant of $\m{A}$ in real CAs is a real quantity,
the condition can be satisfied if and only if $\det (\m{B})\ge 0$.
This is in agreement with explicit formulas for $n\le3$.

\section{Conclusions}\label{sec:conclusions}
First, we have shown analytically  that the  square root of
general MV in $n=p+q\le 3$ Clifford algebras (CA) can be expressed
in radicals and have provided a detailed analysis and algorithm to
accomplish the task. For a general MV the algorithm is rather
complicated, where many of conditions are controlled by plus/minus
signs. In our first paper~\cite{Dargys2019} only the roots of
individual grades have been considered, where  explicit formulas
in a coordinate-free form are provided.

Second, the paper shows that MV roots may be isolated (up to 16
  roots in case of \cl{2}{1}) and/or continuous, or conversely
there may be no roots at all.  Thus, the MV algebras  may also
accommodate a number free parameters that bring in a continuum of
roots on respective parameter hypersurface.

Third,  the described algorithm was implemented in
\textit{Mathematica} system~\cite{AcusDargys2023} and applied in
checking up algorithms  by purely numerical root search. For this
purpose {\it Mathematica} universal root search algorithm was
realized and used in the system function \textbf{FindInstance[~]}
to check whether there are cases when isolated root algorithm
fails. No such cases were found. The only complication we
encountered in the  algorithm  programming was that {\it
Mathematica} symbolic zero detection algorithm
\textbf{PossibleZeroQ[~]} in the more complicated cases often
switched over to numerical procedure to detect that involved symbolic expression with nested radicals
indeed represents zero. This is quite understandable, since it is
well known that two expression equivalence problem is, in general,
undecidable.

Fourth, we found that for  algebras, $\cl{3}{0}$ and $\cl{1}{2}$,
the square root solution in general is a union of the following
sets: 1) when $s^2\neq S^2$, the set consists of (up to) four
different isolated roots, 2) when $s^2 =S^2\neq 0$, the set
consists of two isolated roots and 3) when $s^2 =S^2= 0$
there appears a continuum of roots that belong to four or
smaller dimensional parameter manifolds.  Similar sets with minor
modifications exist for remaining algebras, $\cl{2}{1}$ and
$\cl{0}{3}$ as well.

The proposed algorithm is a step forward in solving general
quadratic equations in CAs (examples are given
in~\cite{Dargys2019}), and may find new applications in the
control and systems theory~\cite{Abou2003}, partly because
presented solutions uncover totally new properties of square root
of MV, for example, the root multiplicity and appearance of free
parameters in the roots. Due to intricacies of square root
algorithms, it is recommended to do all calculations by prepared
in advance  numerical/symbolic subroutines.


\appendix

\section{Square roots in $\cl{1}{0}$ and $\cl{0}{1}$ algebras}\label{appendix:dim1}
In $\cl{1}{0}$ and $\cl{0}{1}$, the square root of general MV
$\m{B}=b_0+b_1 \e{1}$  has the solution $\m{A}=\sqrt{\B}=s+v_1
\e{1}$, where the real coefficients $s$ and $v_1$ are
\begin{align*}
v_1=\frac{b_1}{2 s};\quad&s= \begin{cases}
  \pm\frac{1}{\sqrt{2}}\sqrt{b_0-\sqrt{D}}\quad \textrm{if}\quad b_0-\sqrt{D}> 0 \quad \textrm{and}\quad D\ge 0, \\
  \pm\frac{1}{\sqrt{2}}\sqrt{b_0+\sqrt{D}}\quad \textrm{if}\quad b_0+\sqrt{D}> 0 \quad \textrm{and}\quad D\ge 0,
\end{cases}
\end{align*}
where
\begin{align*}
& D= \begin{cases}
     b_{0}^2 - b_{1}^2, \quad \textrm{for}\ \cl{1}{0},\\
     b_{0}^2 + b_{1}^2, \quad \textrm{for}\ \cl{0}{1}.
   \end{cases}
\end{align*}

When $s=0$, (i.e. when $b_0\pm\sqrt{D} = 0$) and $b_1=0$ the
square roots are
\begin{align*}
&
\m{A}=\begin{cases}
\pm\sqrt{b_0},& \textrm{if}\quad b_1= 0,\quad  \textrm{for}\ \cl{1}{0},\\
\pm\sqrt{-b_0} , & \textrm{if}\quad b_1= 0,\quad   \textrm{for}\
\cl{0}{1}.
\end{cases}
\end{align*}
Note, that the $\cl{0}{1}$ algebra is isomorphic to the algebra of
complex numbers, so we know in advance that any MV in this algebra
has two roots. The MV determinant for this algebra is
positive definite $D= b_{0}^2 + b_{1}^2\ge 0$ and represents the
square of module of a complex number. We shall always
assume that expressions under square roots are non-negative. For
example, in this case the square root can only exist when $D\ge
0$, and either $(b_0-\sqrt{D})\ge 0$ or $(b_0+\sqrt{D})\ge 0$. If
these conditions cannot be satisfied, then  square roots are
absent.

\section{Square roots in $\cl{2}{0}$, $\cl{1}{1}$ and $\cl{0}{2}$ algebras}\label{appendix:dim2}
Square root $\m{A}$ of general MV $\m{B}=b_0+b_1 \e{1}+ b_2 \e{2}
+ b_3 \e{12}$ in  all three algebras is $\m{A}=s+v_1 \e{1} + v_2
\e{2} + S \e{12}$. The coefficients $(s,S)$ are
\begin{align*}
\mkern+10mu&\begin{cases}
  \Bigl(s=\pm\frac{1}{\sqrt{2}}\sqrt{b_0-\sqrt{D}},S=\pm\frac{1}{\sqrt{2}} \frac{b_{3}}{\sqrt{b_0-\sqrt{D}}}\Bigr),\quad
  \textrm{if}\quad b_0-\sqrt{D} > 0 \ \textrm{and}\ D\ge 0, \\
\Bigl(s=\pm\frac{1}{\sqrt{2}}\sqrt{b_0+\sqrt{D}},S=\pm\frac{1}{\sqrt{2}} \frac{b_{3}}{\sqrt{b_0+\sqrt{D}}}\Bigr),\quad \textrm{if}\quad b_0+\sqrt{D} > 0 \ \textrm{and}\ D\ge 0,
\end{cases}
\end{align*}
where the determinant of MV $\m{B}$ is~\cite{Acus2018,Acus2022},
\begin{align*}
& D= \begin{cases}
     b_0^2-b_1^2-b_2^2+b_3^2, \quad \textrm{for}\ \cl{2}{0},\\
     b_0^2-b_1^2+b_2^2-b_3^2, \quad \textrm{for}\ \cl{1}{1},\\
     b_0^2+b_1^2+b_2^2+b_3^2, \quad \textrm{for}\ \cl{0}{2}.
   \end{cases}
\end{align*}

{\it The case $s\neq 0$}. The coefficients $v_1,v_2\in\A$ then are
given by formulas
\begin{align*}
&v_1=\frac{b_1}{2 s},\quad v_2=\frac{b_2}{2 s}.
\end{align*}

{\it The case $s=0$}. When $b_0-\sqrt{D} = 0$, or $b_0-\sqrt{D} =
0$ and $b_1=b_2=b_3=0$, the coefficients $v_1, v_2$ and $S$ are
connected by single equation  $\pm v_1^2\pm v_2^2\pm b_0\pm
S^2=0$. Therefore, one can search the  solution with respect
to any of coefficients $v_1, v_2$ or $S$, and assume that
remaining two  coefficients are the free parameters. For example,
if we solve with respect to $S$, then the square root for each of
algebras is,
\begin{align*}
\mkern+10mu& \m{A}=\begin{cases}
v_1 \e{1} + v_2 \e{2} \pm\sqrt{-b_0+v_1^2+v_2^2} \e{12},\quad  \textrm{for}\ \cl{2}{0},  \quad \textrm{if}\quad b_1=b_2=b_3=0,\\
v_1 \e{1} + v_2 \e{2} \pm\sqrt{b_0-v_1^2+v_2^2} \e{12},\quad\ \  \textrm{for}\ \cl{1}{1}, \quad \textrm{if}\quad b_1=b_2=b_3=0,\\
v_1 \e{1} + v_2 \e{2} \pm\sqrt{-b_0-v_1^2-v_2^2} \e{12},\quad
\textrm{for}\ \cl{0}{2}, \quad \textrm{if}\quad b_1=b_2=b_3=0.
\end{cases}
\end{align*}
Since the coefficient $S$ is real, the roots exist only when the
expressions under square root are positive. The algebra
$\cl{2}{0}$ is isomorphic to $\cl{1}{1}$.

\vspace{3mm} \textit{Example}.\newline The square root of
$\m{B}=6+2 \e{1}+3\e{2}-4\e{12}$ in various 2D algebras:
\begin{align*}
\mkern+10mu& \m{A}=\begin{cases}
\pm\frac{1}{\sqrt{2 (6+\sqrt{39})}}(6+\sqrt{39}+2 \e{1}+3 \e{2}-4 \e{12})&  \cl{2}{0},\\
  \pm\frac{1}{\sqrt{2}} (1+2 \e{1}+3 \e{2}-4 \e{12}) \textrm{ and }
  \pm\frac{1}{\sqrt{22}}(11+2 \e{1}+3 \e{2}-4 \e{12})& \cl{1}{1}, \\
  \pm\frac{1}{\sqrt{2 (6+\sqrt{65})}} (6+\sqrt{65}+2 \e{1}+3 \e{2}-4 \e{12})& \ \cl{0}{2}.
\end{cases}
\end{align*}
Note that in $\cl{1}{1}$ there are four roots.

\section{Algorithm for square root}\label{appendix:algorithm}
\begin{algorithm}[H]
\SetAlgoLined \SetNoFillComment \LinesNotNumbered

\SetKwInput{KwInput}{Input} \SetKwInput{KwOutput}{Output}
\SetKwProg{Sqrt}{Sqrt}{}{} \Sqrt{$(\m{B})$}{
    \KwInput{$\m{B}=b_0+b_1\e{1}+b_2\e{2}+b_3\e{3}+b_{12}\e{12}+b_{13}\e{13}+b_{23}\e{23}+b_{123}\e{123}$}
    \KwOutput{$\m{A}=s+\bv+(S+\bV)I$ with the property $\m{A}^2=\m{B}$, or no solution}
\tcc{\scriptsize Initialization}
$b_{S}=b_{0}^2-b_{1}^2-b_{2}^2-b_{3}^2+b_{12}^2+b_{13}^2+b_{23}^2-b_{123}^2$\;
$b_{I}= 2 b_{3} b_{12}-2 b_{2} b_{13}+2 b_{1} b_{23}-2 b_{0}
b_{123}$\; $D=b_{S}^2+b_{I}^2$\; \tcc{\scriptsize Compute all
$(t_i,T_i)$ pairs} \uIf{$-b_{S}+\sqrt{D}>0$}{$t_{1,2}=\frac{1}{4}
\Bigl(b_{123} \pm \frac{1}{\sqrt{2}}\sqrt{-b_{S}+\sqrt{D}}\Bigr),\
    T_{1,2}=\frac{1}{4}\Bigl(\frac{\pm b_{I}}{\sqrt{2}\sqrt{-b_{S}+\sqrt{D}}}-b_{0}\Bigr)$\;}
    \uElseIf{$-b_{S}+\sqrt{D}=0\ and\ b_{S}>0$}{$t_{1,2}= \frac{1}{4}b_{123}, \
    T_{1,2}= \frac{1}{4} \bigl(\pm\sqrt{b_{S}}-b_{0}\bigr)$\tcc*[r]{\scriptsize degenerate case}}\Else{\Return{$\emptyset$}\tcc*[r]{\scriptsize no root}}
  \tcc{\scriptsize for each $(t_i,T_i)$ pair (where $i=1,2=(+,-)$) find corresponding $(s_i,S_i)$ pair}\vskip 5pt
  \ForEach{$(t_i,T_i)$}{$
  s_{i} =\pm\sqrt{-T_i+\sqrt{T_i^2+ t_i^2}},\qquad S_{i}
  =\pm\frac{t}{\sqrt{-T_i+\sqrt{T_i^2+ t_i^2}}}$\;
  \tcc{\scriptsize every index $i=1,2=(+,-)$ matches two values}
  }
\tcc{\scriptsize For each $(s_i,S_i)$ pair compute corresponding
$v_{i_k},V_{i_k}$}
  \ForEach{$(s_i,S_i)$}{\eIf{$s^2+S^2\neq 0$}{\tiny$
  \begin{aligned}
    v_{i_1}=\frac{b_1s+b_{23}S}{2(s^2+S^2)},&v_{i_2}=\frac{b_{2}s-b_{13}S}{2(s^2+S^2)},& v_{i_3}=\frac{b_3s+b_{12}S}{2(s^2+S^2)}\\
    V_{i_1}=\frac{b_{23}s-b_{1}S}{2(s^2+S^2)},& V_{i_2}=-\frac{b_{13}s+b_{2}S}{2(s^2+S^2)}, & V_{i_3}=\frac{b_{12}s-b_{3}S}{2(s^2+S^2)}
\end{aligned}\;
$\normalsize\newline
  \Return{$\m{A}\leftarrow (s_i,S_i, v_{i_k},V_{i_k})$ }\tcc*[r]{\scriptsize isolated root}}{
    \eIf{$b_1=b_2=b_3=b_{12}=b_{13}=$ $b_{23}=0$\tcc*[r]{\scriptsize $s^2+S^2=0$ case}}{solve any two of $v_{i_k},V_{i_j}$ from \tiny$\begin{cases}b_0=\bv^2-\bV^2,\\ b_{123}=2(\bv\cdot\bV)\end{cases}$\normalsize\newline
      \Return{$\m{A}\leftarrow (s_i,S_i, v_{i_k},V_{i_j})$ }\tcc*[r]{\scriptsize continuum of roots}}{
    \Return{$\m{A}\leftarrow\emptyset$ }\tcc*[r]{\scriptsize no root}
      }
  }}
} \label{AlgForSqrt30} \caption{Algorithm for square root for $\cl{3}{0}$.}
\end{algorithm}

The algorithm in~\ref{appendix:algorithm} was implemented by {\it
Mathematica}~\cite{AcusDargys2023} using single piecewise function
{\bf Piecewise[~]}  to which  special case solutions then were
added in the final output. Such an approach also allows to compute
isolated square roots of MV with symbolic coefficients.
Algorithms for remaining $n=3$ algebras are similar to
\cl{3}{0} in~\ref{appendix:algorithm}. We have found the
algorithms very helpful in checking over and dealing with real MV
roots.

\subsection*{Acknowledgment}
This research was partly (A.~Acus) funded by the European Social
Fund under Grant No. 09.3.3-LMT-K-712-01-0051. Authors want to
thank  Vanessa Hollmeier for detected flaw in the Appendix of
preprint version~1 (removed in this version) about square root of
matrix.




\end{document}